# Analog Codes on Graphs

**Nandakishore Santhi**
nsanthi@ucsd.edu

**Alexander Vardy**
vardy@kilimanjaro.ucsd.edu

Department of Electrical and Computer Engineering,
University of California San Diego,
9500 Gilman Drive, MC 0407, La Jolla, CA-92093-0407, USA.

**Abstract**

We consider the problem of transmission of a sequence of real data produced by a Nyquist sampled band-limited analog source over a band-limited analog channel, which introduces an additive white Gaussian noise. An analog coding scheme is described, which can achieve a mean-squared error distortion proportional to $(1 + SNR)^{-B}$ for a bandwidth expansion factor of $B/R$, where $0 < R < 1$ is the rate of individual component binary codes used in the construction and $B \geqslant 1$ is an integer. Thus, over a wide range of SNR values, the proposed code performs much better than any single previously known analog coding system.

## I. Introduction

We consider the problem of transmission of a sequence of real data produced by a band-limited analog source over a band-limited analog channel, which introduces an additive white Gaussian noise. A similar setting arises in several practical situations. A traditional approach towards solving this problem is through the application of Shannon's source-channel separation principle. This involves the separate design of optimal (digital) source and channel codes, and using these codes in *tandem* [MP02]. However by invoking the separation principle, we sacrifice one of the most important advantages of an analog communication system, namely its ability to perform well under varying noise levels. In literature, this property has been referred to as "robustness". By contrast, a digital communication system employing separate source and channel codes typically can operate satisfactorily only within a very narrow region around the *designed* signal to noise ratio. Another important property of an analog system is the graceful degradation in performance with an increasing noise level. This property is very valuable for a good broadcast system. However analog systems have several disadvantages too. In practice an analog system is complicated to design and maintain, is susceptible to drift with time, and expensive. In addition, practical systems thus far suffer from the *threshold-effect* – beyond a certain signal-to-noise ratio, the system performance tends to either saturate or improve only marginally with more transmit power. In spite of several such shortcomings, analog systems are still rated high in applications such as transmission and recording of music, due to the lack of granularity which is the result of imperfect source quantization.

In order to compare various analog systems, it is customary to measure the end to end distortion in terms of the mean squared error. In an effort to explain the threshold phenomena in analog communication systems, Ziv [Ziv70], considered a broad class of well behaved modulation signals which include all currently employed analog schemes. He showed that for this class with bandwidth expansion factors over unity, as the SNR $\gamma$ increases, the mean squared distortion cannot fall at a rate faster than $\gamma^{-2}$. Shamai, Verdú and Zamir [SVZ98] demonstrated a joint source-channel coding scheme that is optimal for a class of sources and channels. In [MP02], Mittal and Phamdo presented several hybrid digital-analog joint source-channel codes which are shown to be "nearly robust" for the case of broadcast with two receivers. For the case of broadcast to two receivers, a distortion pair $(D_1, D_2)$ is said to be achievable if the

This work was supported in part by the National Science Foundation.

first listener achieves $D_1$ and the second listener $D_2$ simultaneously. In a recent work Reznic, Feder and Zamir [RFZ05] consider the problem of analog broadcast to two listeners at bandwidth expansion ratios of more than 1. They prove a lower bound for the achievable rate distortion pairs using information theoretic arguments. Using some necessary conditions, they show that if a system is optimal at a certain high enough $\gamma$, then its distortion cannot decay at a rate faster than $\gamma^{-1}$. Thus one of the Mittal-Phamdo schemes is seen to be optimal at high signal to noise ratios.

In addition to the hybrid code constructions due to Shamai et al [SVZ98], Mittal et al [MP02], and Reznic et al [RFZ05], there have been attempts to construct purely analog coding schemes for the general analog channel with bandwidth expansion factors of larger than unity. Chen and Wornel [CW98] give an analog code construction based on chaotic maps for uniformly distributed analog source, along with an efficient decoding algorithm using estimation techniques. Vaishampayan and Costa [VC03] construct analog codes using linear dynamical systems and provide efficient decoding algorithms. Using curves on higher dimensional spheres and topological arguments they show that their codes have much in common with the Chen et al chaos map codes. In both the above constructions, the mean square distortion cannot decay at a rate better than $\gamma^{-3/2}$.

All these prior results seem to indicate that strong analog coding schemes invariably suffer from the analog threshold effect, further limiting their practical usefulness. Shannon's rate distortion theory on the other hand suggests that mean square distortion may fall at a rate of $\gamma^{-N}$ at a bandwidth expansion factor of $N$. It has been widely conjectured that no single analog (or hybrid analog-digital) scheme can achieve this rate of fall with SNR. In this paper, we show that this is not necessarily so. Here we construct a family of analog codes, using digital component codes. The component codes are used in a "power-splitting" superposition scheme with an infinite number of levels. This is similar in principle to the superposition scheme introduced by Cover [Cov72] and related to the multilevel codes considered by Calderbank and Seshadri in [CS93]. Assuming unbounded complexity is admissible at both the transmitter and receiver, we analyze the performance of these codes on an AWGN channel. We then consider some practical component codes and show that under most practical situations, the codes constructed herein can give a rate of fall of mean-square-distortion well above $\gamma^{-2}$ over a wide range of SNRs. It must however be noted that our results do not contradict the conclusions of either Ziv [Ziv70] or Reznic et al [RFZ05]. Instead the class of codes constructed here do not fall under the type of modulation signals considered by Ziv. The analog codes described here achieve a fall in distortion proportional to $\gamma^{-B}$, where $B$ is an integer such that $1 \leqslant B \leqslant \lfloor R \cdot N \rfloor$, where $N$ is the bandwidth expansion factor and $0 < R < 1$. Moreover though it never achieves any distortion point which is on the lower bound given by the rate-distortion function, the achievable MSE distortion can be made to fall almost parallel to the lower bound at high enough SNR.

In the next section, we give the construction of our code, and several practical examples. We also show why the codes we construct do not fall into the class of codes considered by Ziv. In the third section, we analyze the performance of the proposed system assuming use of capacity achieving component binary codes. We also give a distortion upper bound using the union-bounding technique for hard decision maximum likelihood decoding for the component binary codes. We give simulation results for analog codes constructed using repeat-accumulate codes as the component codes. The analog decoder used in this case employs the sum-product algorithm decoder along with a hierarchical successive cancellation technique.

## II. Theoretical Bound from Rate-Distortion Theory

We consider a discrete time analog source which draws values independently and uniformly from the interval $\mathcal{S} = [-\sqrt{3}, \sqrt{3})$. The average source power is thus constrained to be unity. We may assume that, such a discrete time memoryless source which outputs at a rate of $2W_S$ samples per second can be constructed from a band-limited continuous time source by means of appropriate sampling. Let us denote by $\mathbf{s} = \{s_t\}$, the sequence of continuous valued random symbols generated by this source.

We wish to transmit the source output to listeners across a band-limited and power-limited continuous time analog channel which adds a white Gaussian noise to the input signal. The channel can also be represented by an equivalent discrete time version using the Nyquist sampling theorem. The channel is used at the rate of $2W_C$ per second. Thus if the channel input, output and noise are denoted by $y_t$, $r_t$ and $v_t$ respectively, then they are related as $r_t = y_t + v_t$ at time instant $t$. Without loss of generality, the channel input is assumed to be subject to the input power constraint $E_W[y_t^2] = 1$.

By rate-distortion theory, for a memoryless uniform source, the rate distortion function can be obtained as a lower-bound on a difference in differential entropy:

$$R(D) = h(s) - h(s|\hat{s}) = -h(s|\hat{s}) \geqslant \tfrac{1}{2}\log_2\left(\tfrac{1}{2\pi e D}\right) \text{ (bits/use)} = W_S \log_2\left(\tfrac{1}{2\pi e D}\right) \text{ (bits/s)}$$

Similarly, the capacity of the discrete-time AWGN channel is:

$$C(\sigma^2) = \tfrac{1}{2}\log_2\left(1 + \tfrac{1}{\sigma^2}\right) \text{ (bits/use)} = W_C \log_2\left(1 + \tfrac{1}{\sigma^2}\right) \text{ (bits/s)}$$

The two relations together give the following lower bound on the achievable mean squared-error distortion for analog transmission of a uniform source over the discrete memoryless AWGN channel:

$$D(\sigma^2) \geqslant \frac{1}{2\pi e(1 + \frac{1}{\sigma^2})^N}$$

where, $N = W_C/W_S$ is called the *bandwidth expansion factor*. We call $1/\sigma^2$ the *signal to noise (power) ratio*.

An *analog encoder* of length $n$ is a block encoder which maps the real source samples grouped $k$ at a time to a vector of real values of length $n$. Thus the encoder $\mathcal{E}_k$ is a mapping $\mathcal{E}_k : \mathrm{R}^k \mapsto \mathrm{R}^n$, such that the encoded sequence satisfies the input power constraint of the channel, $\frac{1}{n}E_S[||\mathcal{E}_k(S)||^2] \leqslant 1$, where $S$ represents a block of $k$ source samples. Similarly, the *analog decoder* is a mapping $\mathcal{D}_k : \mathrm{R}^n \mapsto \mathrm{R}^k$. The received signal is $Z = W + V$ where $W = \mathcal{E}_k(S)$. A family of analog codes, is a sequence of encoder-decoder pairs with increasing block-length $k$, and a fixed “rate” $R \triangleq \frac{k}{n}$. Our objective is to construct a family of analog codes which achieve sufficiently low mean squared error distortion in the limit of very large block lengths, $D(\sigma^2) = \lim_{k\to\infty} \frac{1}{k}E_S[||S - \widehat{S}||^2]$, where $\widehat{S}$ is the vector of estimates produced by the decoder $\mathcal{D}_k$. Below we give our code construction.

## III. Construction of the Analog Codes

Let a realization of the source output at time $t$ be denoted as $s_t$. Since the source has mean zero and unit variance uniform pdf, scale the realization to obtain a uniform distribution in the interval $[0, 1)$ by forming $x_t = (s_t + \sqrt{3})/(2\sqrt{3})$. Now we represent $x_t$ by its terminating binary representation. Let this binary representation be given by $\mathbf{X}_t = \{X_{1,t}, X_{2,t}, \cdots, X_{\ell,t}, \cdots\}$. It is well known [Res98] that, $X_{\ell,t}$ are Bernoulli random variables with $p = 1/2$. For a fixed integer parameter $B$, form for each $i \in \{1, 2, \cdots\}$ sequences of the form

$$\mathbf{U}_i = \{\ \cdots,\ X_{t-1,iB}, X_{t,(i-1)B+1}, X_{t,(i-1)B+2},\ \cdots,\ \ X_{t,iB}, X_{t+1,(i-1)B+1}, X_{t+1,(i-1)B+2},\ \cdots,\ \}$$

Now using *component binary codes*, encode the sequences $\mathbf{U}_i$ for each $i$ separately. Denote by $\mathbf{W}_i$ the resulting sequence of Bernoulli random variables. Each such $i$ is referred to as the $i^{th}$ *encoded (bit) level*. By a suitable choice of the component code, it is always possible to ensure that this sequence is composed of Bernoulli random variables that are distributed fairly, with $p = 1/2$. The encoding of the reordered source sample bit planes is shown in Figure 1.

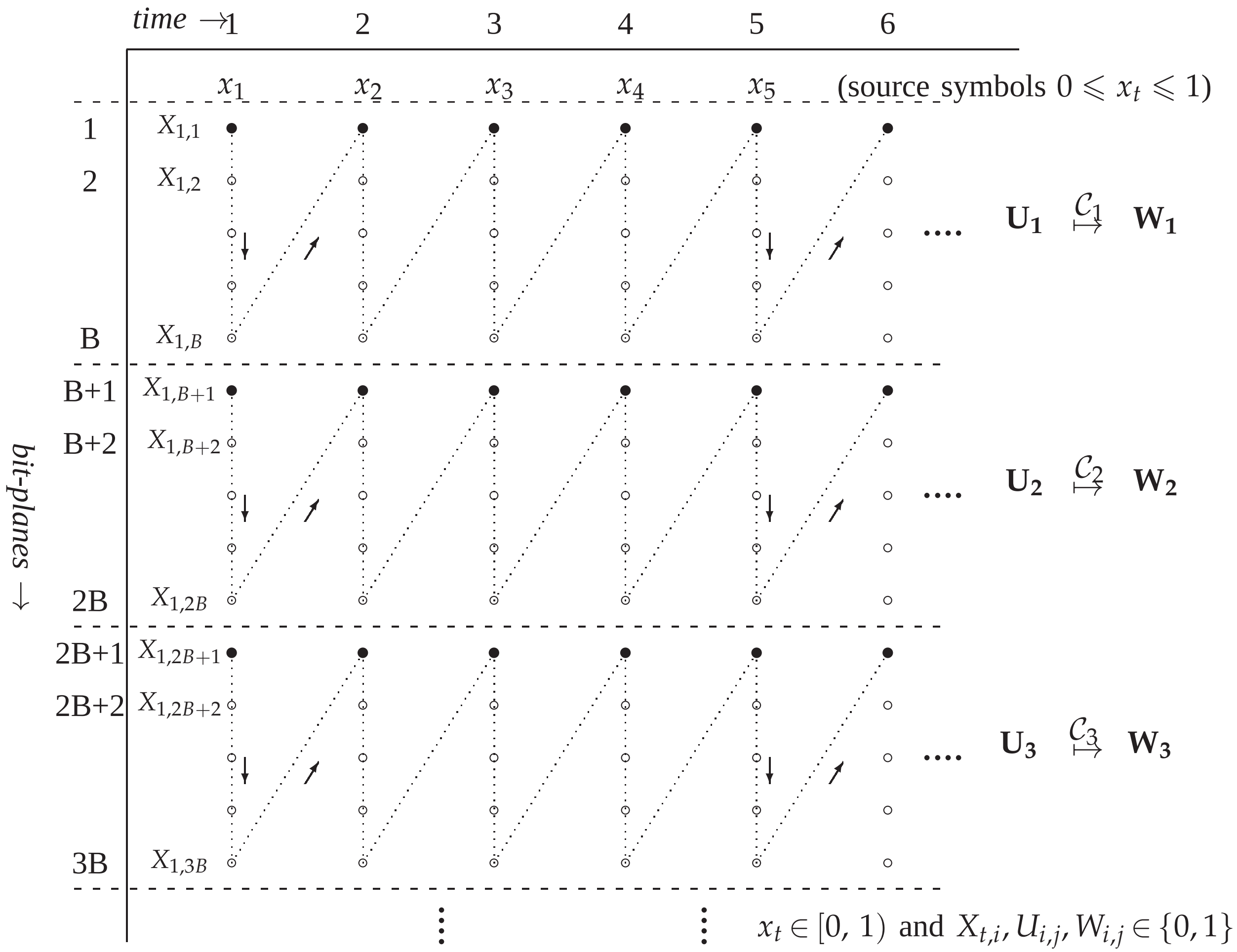

**Figure 1** *The source bit planes are reordered and encoded using component binary codes $\mathcal{C}_i$ at each level $i$.*

Now define a mapping $f_w : \{0,1\}^* \to \mathrm{R}$ for a binary sequence $\mathbf{b} \in \{0,1\}^*$ as,

$$f_w(\mathbf{b}) \triangleq \sum_{i=1}^{\infty} (2b_i - 1) \cdot w_i \tag{1}$$

where, $w_i = w^{\frac{1}{2}} \cdot (1-w)^{\frac{i-1}{2}}$ for some $w \in [0,1)$. This mapping will be often referred to as the *analog encoding map*.

The encoded sequence of bits $\mathbf{W}_i$ are now read off vertically as shown in Figure 2 and the analog mapping function $f_w(.)$ is applied with some fixed weight, $w$. The real valued vector $\mathbf{y}$ so obtained is the analog code corresponding to the source vector $\mathbf{s}$. The analog encoding map satisfies the unity input power constraint for the analog channel and the resulting vector is transmitted over the channel.

As an example, let us take the simplest case, when $B = 1$. Then, the above code construction would be equivalent to encoding each one of the $i^{th}$ bit planes at significant positions $i$ separately, followed by the mapping $f_w(\cdot)$ which is invoked at each time instant. The analog encoder block diagram is depicted in Figure 3.

## IV. The main results

The following lemma brings out the self similar, scaled nature of the pdf of the real random variables, $y_j$ produced by the code construction as given in the previous section.

**Lemma IV.1** *Let* $\mathrm{W} = \{W_1, W_2, \cdots, W_n, \cdots\}$ *be a sequence of independent identically distributed Bernoulli random variables taking on values* $\{0,1\}$ *with* $p = 1/2$. *Define the map* $f_w$ *as in (1). Then, the random variable* $\mathbf{y} = f_w(\mathrm{W})$ *has the following properties:*

*(i)* $E[\mathbf{y}] = 0$

*(ii)* $Var[\mathbf{y}] = E[\mathbf{y}^2] = 1$

*(iii)* $\mathbf{y}$ *has a fractal probability measure which draws values from the interval* $[-\frac{\sqrt{w}}{(1-\sqrt{1-w})}, \frac{\sqrt{w}}{(1-\sqrt{1-w})})$.

*(iv) if* $w = 3/4$, *then* $\mathbf{y}$ *is distributed uniformly in* $[-\sqrt{3}, \sqrt{3})$.

*Proof.*

(i) This follows from the fact that the independent Bernoulli trials are fair.

(ii) Denote $\alpha \triangleq \sqrt{w/(1-w)}$ and $\beta \triangleq \sqrt{(1-w)}$. Then, $w_i = \alpha \cdot \beta^i$. Now, observe that $E[\mathbf{y}^2] = \sum_{i=1}^{\infty} E_{x_i}[(2W_i - 1)^2] \cdot w_i^2 = \alpha^2 \cdot \sum_{i=1}^{\infty} \beta^{2i} = \alpha^2 \cdot \beta^2/(1-\beta^2) = 1$. The contribution to the variance due to the variable $X_1$ is $w$.

(iii) The interval for $\mathbf{y}$ extends symmetrically to a length of $\sum_{i=1}^{\infty} w_i = \alpha \cdot \sum_{i=1}^{\infty} \beta^i = \alpha \cdot \beta/(1-\beta) = \sqrt{w}/(1-\sqrt{1-w})$ about the origin. Now, consider the sequence, $\mathrm{W}^* = \{W_2, W_3, \cdots, W_n, \cdots\}$ which can be formed from the sequence $\mathrm{W}$ by omitting the first random variable $W_1$. Clearly, the two random variables, $\mathbf{y}^* = f_w(\mathrm{W}^*)$ and $\mathbf{y} = f_w(\mathrm{W})$ have the same probability density function, because $W_i$ are iid giving, $p_{(\mathbf{y}^*,w)} = p_{(\mathbf{y},w)}$. But the probability density function $p_{(\mathbf{y},w)}$ can also be expressed in terms of $p_{(\mathbf{y}^*,w)}$ in a different way. To see this, observe that $W_1 = 0$ or 1 with equal probability, and hence, $\mathbf{y} = \beta(\mathbf{y}^* \pm \alpha)$ with equal chance. From these we get, $p_{(\mathbf{y}^*,w)}(y) = p_{(\mathbf{y},w)}(y) = (p_{(\mathbf{y},w)}(y/\beta - \alpha) + p_{(\mathbf{y},w)}(y/\beta + \alpha))/(2\beta)$. This shows that, the pdf of $\mathbf{y}$ is formed from scaled and translated versions of itself. See Figure 4 for an example with $w = 0.05$.

(iv) For the case $w = 3/4$, $\mathbf{y} \in [-\sqrt{3}, \sqrt{3})$ from (iii). To prove that $\mathbf{y} \sim U([-\sqrt{3}, \sqrt{3}))$, take $\mathrm{W}$ to be the terminating binary representation of $\mathbf{x} = (\mathbf{z} + \sqrt{3})/2\sqrt{3}$ where, $\mathbf{z} \sim U([-\sqrt{3}, \sqrt{3}))$, then by construction, $\mathbf{y} = \mathbf{z}$ for all realizations. ∎

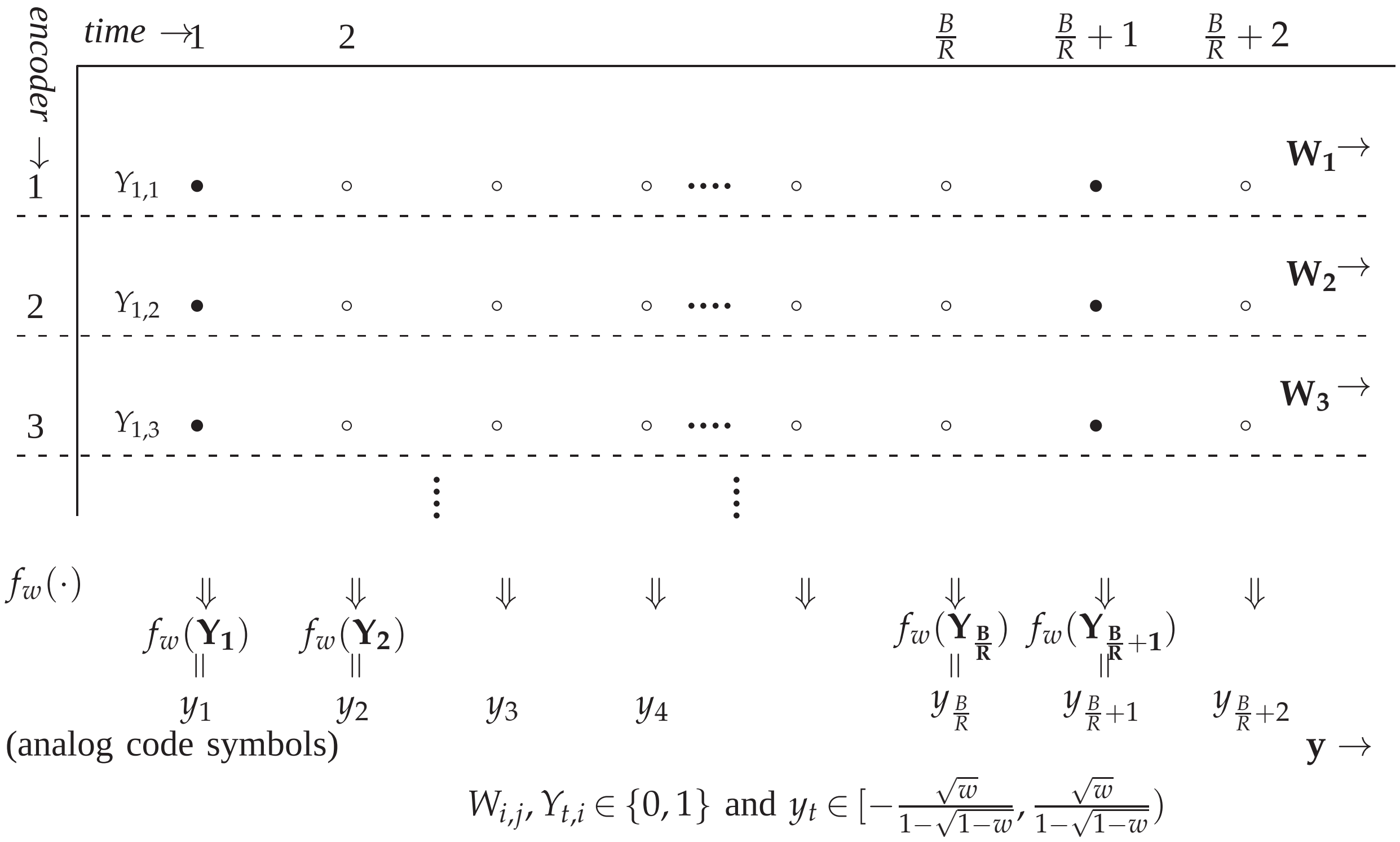

**Figure 2** *The encoded bit stream is mapped to the analog channel code using the function $f_w(.)$ at each instant and across all encoder levels. The output is a sequence of real valued random variables which unit variance.*

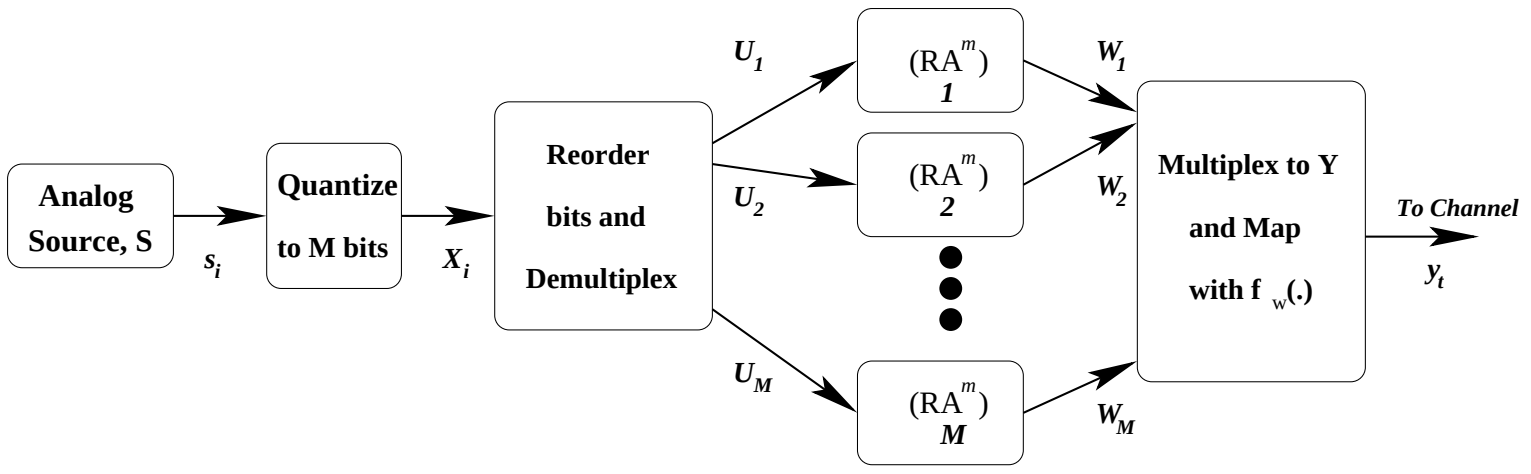

**Figure 3** *Block diagram of the Encoder which maps the Analog source output into the code symbols in* $\mathbb{R}$. *Where as the code construction is valid for any countable integer* $M$*, due to practical considerations, one may have to limit implementations to a finite value of* $M$.

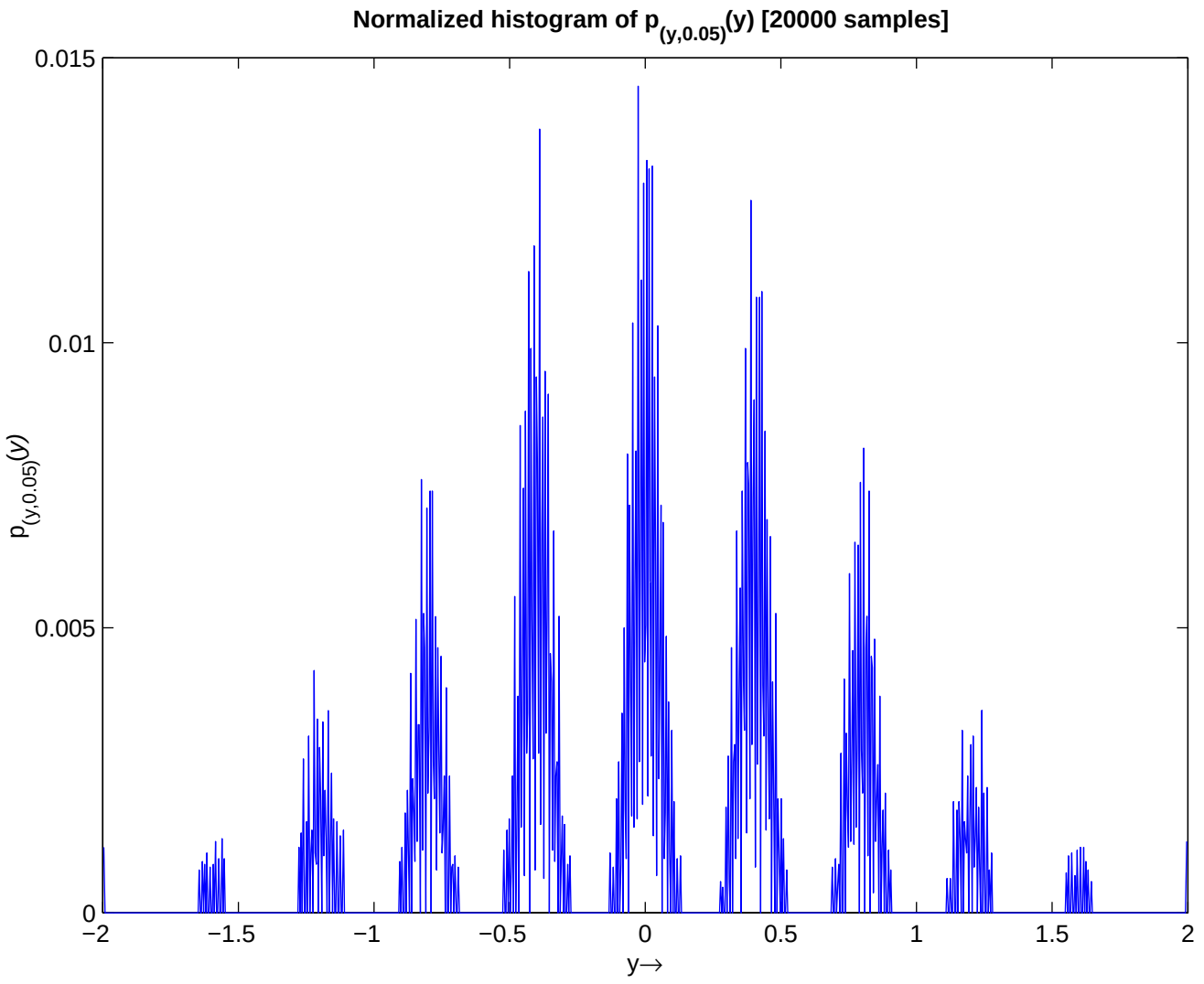

**Figure 4** *Histogram showing* $p_{(y,w=0.05)}(y)$*, with* 20000 *samples: Self similarity under scaling is apparent.*

We now fix a few notations to be used later. Let us denote by $C(p_{\mathbf{n}_\gamma}, \gamma)$, the Shannon channel capacity of a soft output, additive noise channel with a noise pdf of $p_{\mathbf{n}_\gamma}(n)$, at an output SNR of $\gamma$, when the input alphabet is restricted to be binary. We also denote by $p_{(\mathbf{y},w)}$, the pdf of $\mathbf{y} = f_w(\mathrm{Y})$. Define a new random variable, $\mathbf{z}_{\gamma,w} = \beta\mathbf{y} + \mathbf{n}_\gamma$, where as before $\beta = \sqrt{(1-w)}$. If the random processes $\mathbf{y}$ and $\mathbf{n}_\gamma$ are independent, then the three pdfs are related as, $p_{\mathbf{z}_{\gamma,w}}(z) = p_{\mathbf{n}_\gamma}(n) \circledast p_{(\mathbf{y},w)}(y/\beta)/\beta$, with $\circledast$ denoting a linear convolution.

The bandwidth expansion and the minimum MSE distortion achieved for the analog codes constructed as in Section III are related. The scaled nature of the pdf of the noise contributed by the less significant bits of the binary representation of $y_j$ leads us to a characterization of the performance of the analog codes.

**Theorem IV.1** *Let a real analog source* $S$ *draw independently with a uniform distribution from the interval,* $\mathcal{S} = [-\sqrt{3}, \sqrt{3})$*, so that the mean source power is restricted to be unity. Let the source outputs be transmitted using a real alphabet, additive noise channel, with a noise pdf of* $p_{\mathbf{n}_\gamma}(n)$*. Also let the source process be independent of the channel noise process. Then for any integer* $B \geqslant 1$*, at a bandwidth expansion factor* $B/C(p_{\mathbf{z}_{\gamma,w}}, w\gamma/(1+(1-w)\gamma))$ *it is possible to simultaneously achieve for different integers* $k \geqslant 1$*, MSE distortions of* $(1-w)^{k \cdot B}$ *at corresponding SNR of* $\gamma/(1-w)^{k-1}$*, using capacity achieving channel codes designed for an additive noise memoryless channel with noise pdf given by* $\mathbf{z}_{\gamma,w}$.

*Proof.*

The proof makes use of the explicit code construction presented in Section III. The binary channel encoders used in the construction are chosen, so that they achieve the capacity $C(p_{\mathbf{z}_{\gamma,w}}, w\gamma/(1+(1-w)\gamma))$ over the channel with additive noise of pdf $p_{\mathbf{z}_{\gamma,w}}(z)$. Recall that, the most significant bit at instant $j$ of the transmitted real symbol $y_j$ was denoted by $y_{0,j}$. The bit $y_{0,j}$ sees a channel which is equivalent to one with an additive noise of pdf $p_{\mathbf{z}_{\gamma,w}}(z)$ and an output SNR of $w\gamma/(1+(1-w)\gamma)$, as shown in Figure 5. At the receiver, the received value will be $\hat{\mathbf{r}}_j = W_{0,j} + \mathbf{z}_{\gamma,w}$, which can be decoded to form the original sequence $\mathbf{U}_0$ with an arbitrarily small probability of error, due to the capacity achieving channel coding. This in turn recovers the first $B$ bit planes of the source. The MSE distortion is then only due to the rest of the bits numbered from $B+1$ in the binary representation $X_t$ of the source symbol $s_t$. This is $(1-w)^B$ as claimed. The bandwidth expansion factor is clearly $B/C(p_{\mathbf{z}_{\gamma,w}}, w\gamma/(1+(1-w)\gamma))$.

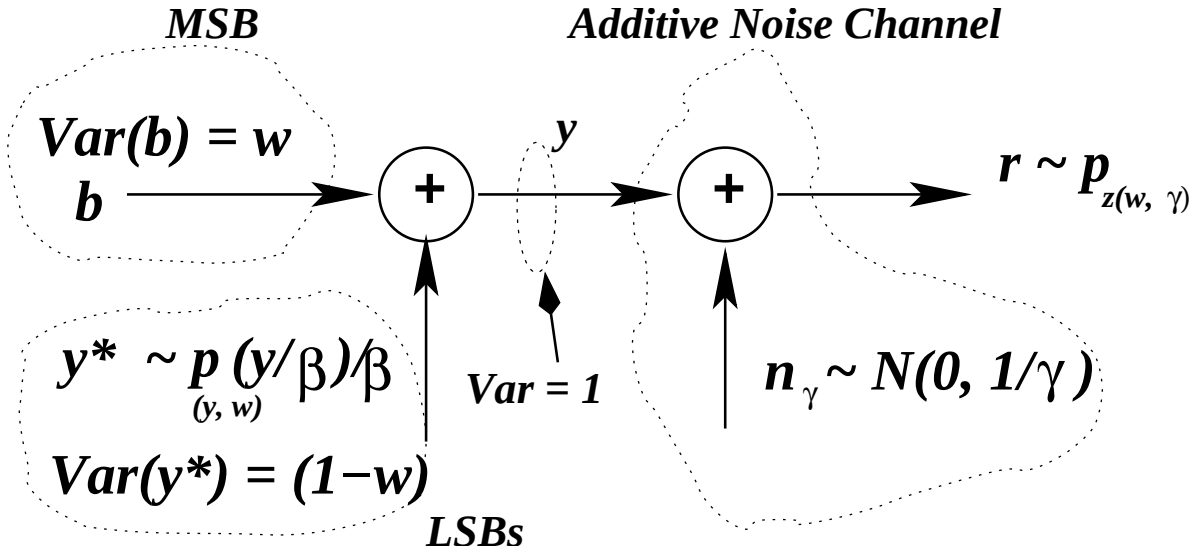

**Figure 5** *The presence of weighted less significant coded bits has the effect of an added noise of variance $(1-w)$ in addition to the AWGN $n_\gamma \sim \mathcal{N}(0, 1/\gamma)$. Here, $b$ is a Bernoulli($p = 0.5$) random variable taking values from $\{\pm\sqrt{w}\}$; $y^*$ has zero mean, variance $(1-w)$, and a pdf given by $p_{(y,w)}(y/\beta)/\beta$. From the figure, the actual SNR for the binary symbol $b$ is, $\gamma' = w\gamma/(1+(1-w)\gamma) \leqslant \gamma$.*

From above discussion, for an output SNR greater than $\gamma$, a distortion of $(1-w)^B$ is achievable for a bandwidth expansion factor of $N = B/C(p_{\mathbf{z}_{\gamma,w}}, w\gamma/(1+(1-w)\gamma))$. This code recovers without error, the binary sequence $\mathbf{U}_1$. From this sequence, we can recover by reordering, the binary sequence representation $\mathbf{X}_j$ of each symbol of the source sequence, $s_j$ correct up to the first $B$ bits(denoted as $\widehat{s}_{j,B}$). The perfectly recovered sequence of bits, $\mathbf{U}_1 = \{U_1\}$ can be re-encoded at the receiver and subtracted from the received sequence, $\mathbf{r} \triangleq \{r_j = W_{1,j} + \mathbf{z}_{\gamma,w}\}$, and then scaled by $1/\beta$, to obtain a new random variable sequence, $\mathbf{r}_1 \triangleq \{r_{1,j} = W_{2,j} + \mathbf{z}'_{\gamma/(1-w),w}\}$ by Lemma IV.1. Invoking the decoder again on $\mathbf{r}_1$, and since $\gamma/(1-w)^k \geqslant \gamma/(1-w)$, we can recover $W_{2,j}$ perfectly. We now proceed by induction on the number of stages of the decoder operation, denoted by $k$. ∎

Let $S$ be a source drawing from an alphabet, $\mathcal{S}$ with a probability distribution, $p_S(s)$ which allows successive refinement of information [EC91] for a distortion function $d$, then similar results may also hold for such sources. Figure 6 gives a block diagram of the receiver structure proposed in the Theorem IV.1.

The $D_2$ versus $\gamma$ curve given by the Theorem IV.1 falls at a rate proportional to $\gamma^{-B}$. In the limit as the rate of the binary code approaches unity, $B \to N$, and the distortion falls at a rate proportional to $\gamma^{-N}$ in the limit. This means that at high enough SNR, the analog code performance curve can be made parallel to the optimal distortion curve for the power limited AWGN channel, given by, $D_2 \geqslant \frac{1}{2\pi e}(1+SNR)^{-N}$. Therefore the analog code presented in this paper can be expected to outperform some other schemes suggested in literature [CW98, VC03] at large $\gamma$.

### *A. Discussion and Example*

In Figure 7, the capacity curve for a binary input channel with two independent additive noise components, one a uniform noise distributed in the interval $[-\sqrt{3}/2, \sqrt{3}/2)$ and the other a Gaussian noise of zero mean and unit variance is shown. In Figure 8 a typical achievable information rate and corresponding SNR are shown. The noise variance at which a rate of $B/\rho$ is achieved on this channel is denoted by $\sigma_*^2$. We now restrict our discussion to the simplest case when $w = 3/4$. In this case, below a noise variance of $\sigma_*^2$, the first level of encoded bit stream can be decoded with arbitrarily small error probability when the bandwidth expansion ratio $N$ is at least $\rho$. Similarly, the second level of encoded bit stream can be decoded with vanishing error probability when the noise variance is below $\sigma_*^2/4$. In general, the $i^{th}$ level can be decoded when the noise variance is below $\sigma_*^2/2^{i-1}$.

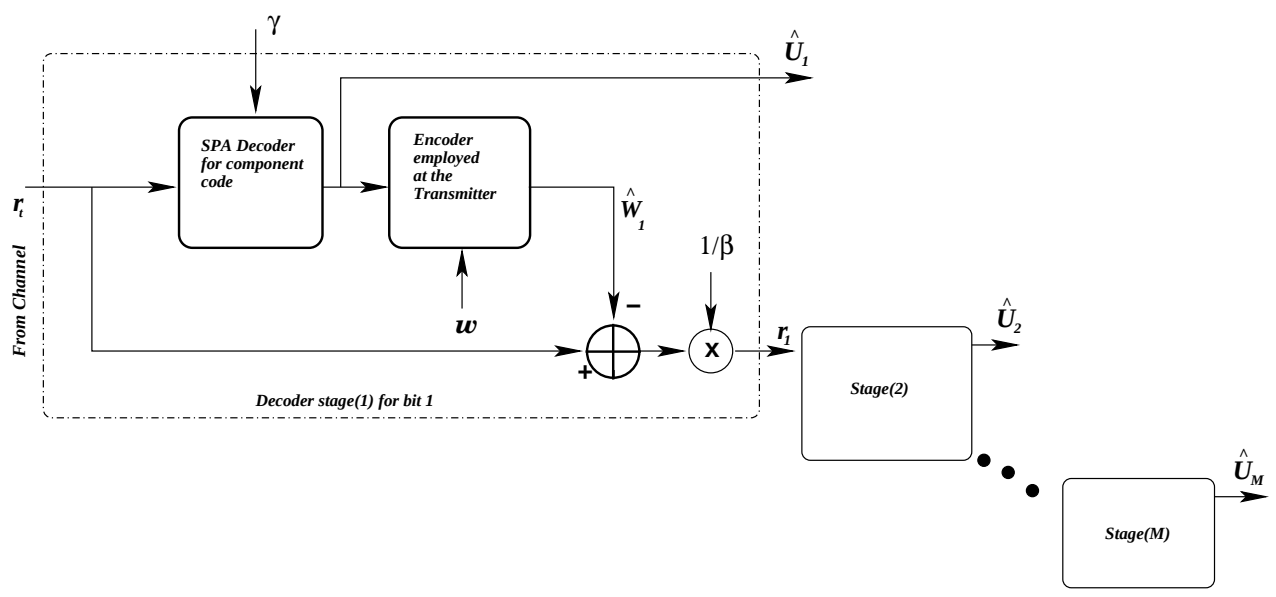

**Figure 6** *The decoder for the Analog Code, with $M$ decoding stages corresponding to the $M$ most significant bits. In the figure, $r_t$ denotes the real valued channel output at time $t$, while $\widehat{U_j}$ and $\widehat{W_j}$ denote the estimates for $U_j$ and $W_j$ (see Figure 3) respectively at the receiver. These estimates $\widehat{U_j}$ can be reordered and used to recover an estimate of the analog source output in an obvious manner.*

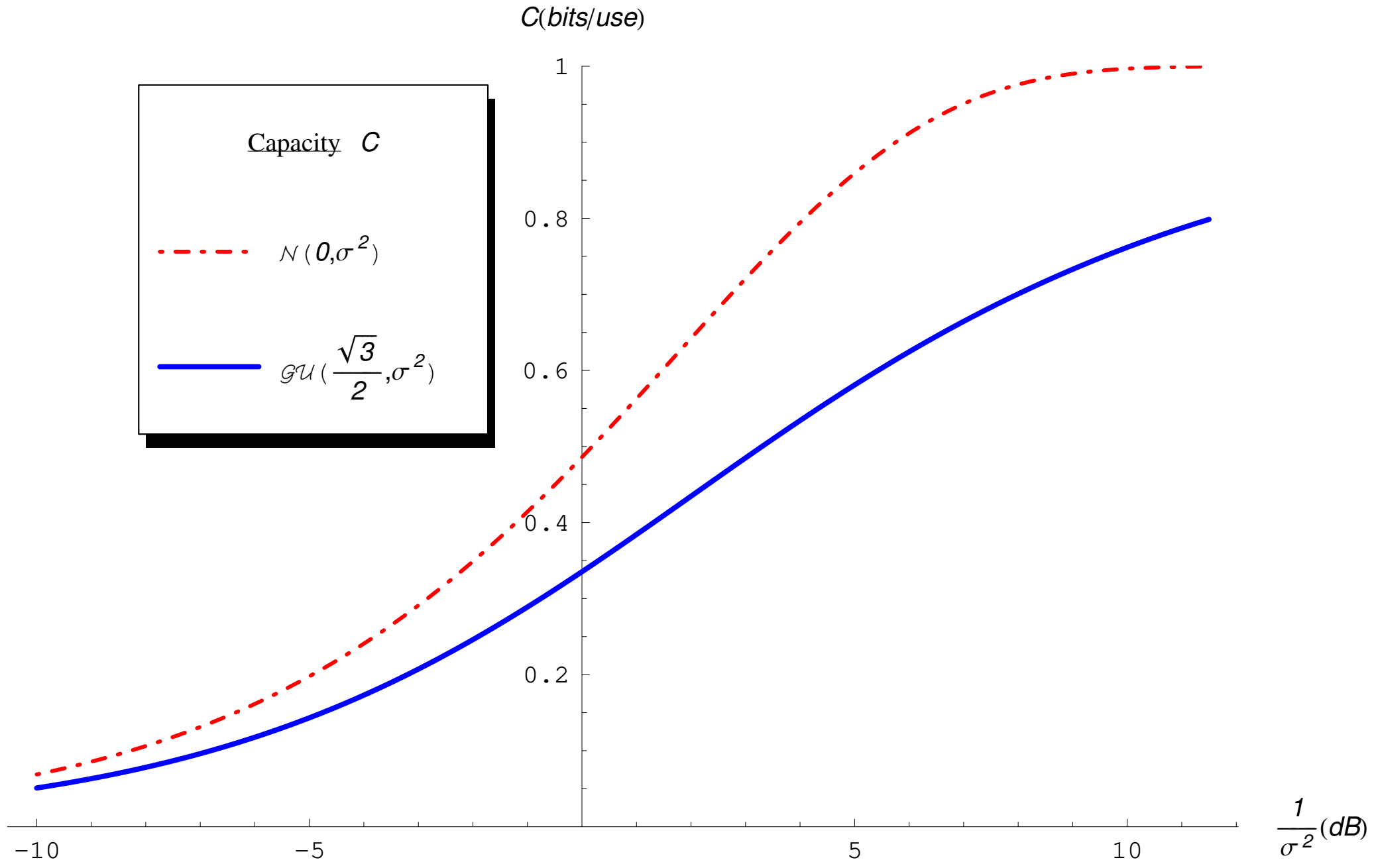

**Figure 7** *The capacity of the two additive noise binary input, real output channels. One of the channels is the AWGN channel with normal pdf $\mathcal{N}(0,\sigma^2)$. The second channel has a Gauss-Uniform pdf $\mathcal{GU}(\frac{\sqrt{3}}{2},\sigma^2)$.*

When $w = 3/4$, the additive noise due to the lower significant bit planes suffered by the bits in the first level of encoded bits is distributed uniformly in the interval $[-\sqrt{3}/2, \sqrt{3}/2)$. The combined pdf of the additive noise presented to the $i^{th}$ level of encoded bits is the convolution of the uniform pdf in the interval $[-a_i, a_i)$ with the Gaussian pdf $\mathcal{N}(0,\sigma^2)$, where $a_i = \frac{\sqrt{3}}{2^i}$. In the remaining sections, we will call

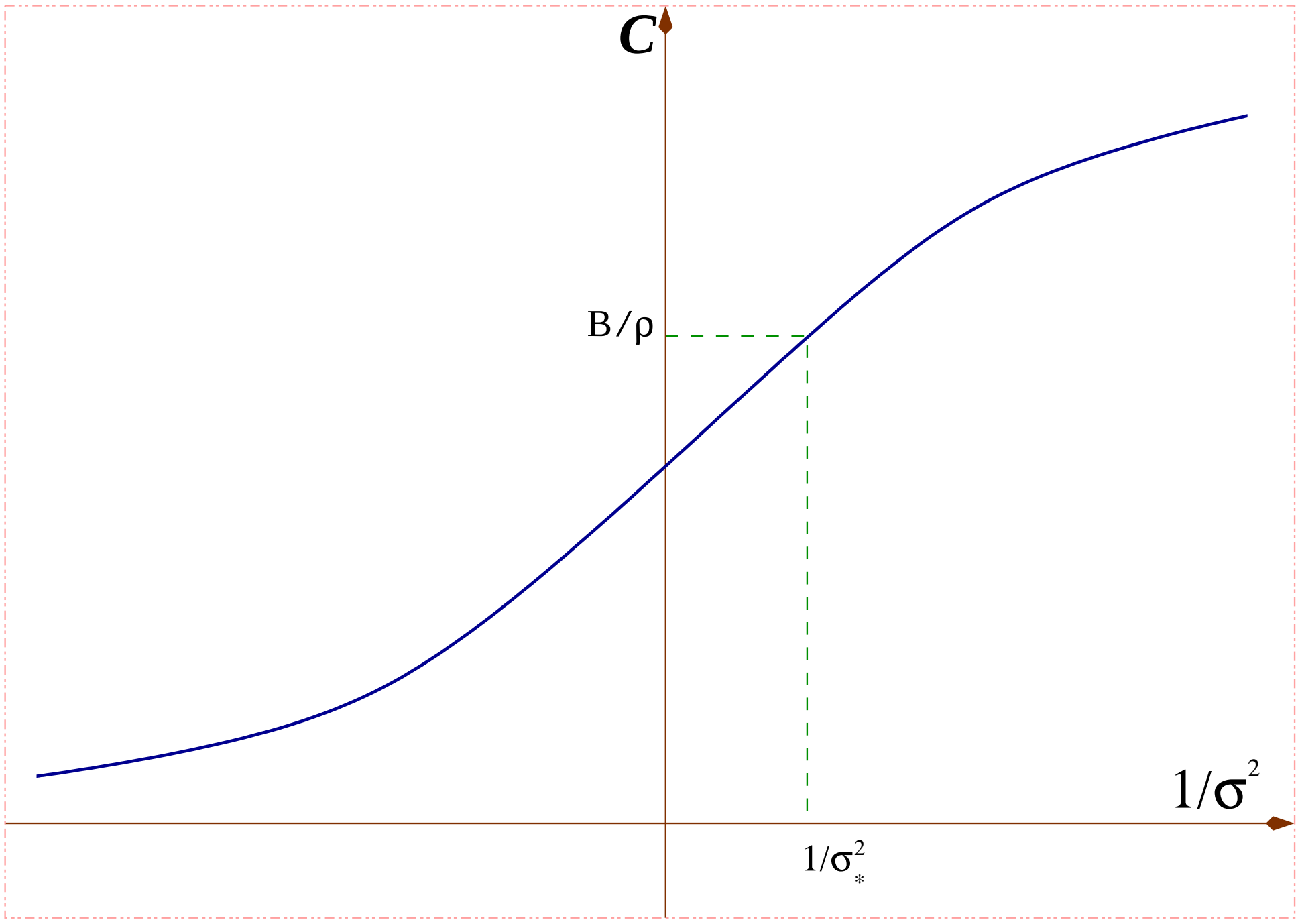

**Figure 8** *The capacity of the Gauss-Uniform noise channel with pdf $\mathcal{GU}(\frac{\sqrt{3}}{2}, \sigma^2)$. A typical achievable rate point based on the ratio $B/\rho$ and the corresponding SNR, $1/\sigma_*^2$, where $N = \rho$ is a particular bandwidth expansion factor.*

this new pdf the *Gauss-Uniform* pdf, and denote it by $\mathcal{GU}(a_i, \sigma^2)$.

## V. Analysis of Finite Length Practical Encoders and Hard Decision ML Component Decoders

We choose the weight factor to be $w = \frac{3}{4}$. Consider the case of $B = 1$ and let all component binary codes be the same. According to the analog code construction outlined earlier, the different bit planes of the source output are encoded separately using binary codes of rate $R$, and then the map $f_w(.)$ is applied to obtain the analog code. The bandwidth expansion factor is then $N = 1/R$. However the guaranteed rate of fall of distortion is only $(1 + SNR)^{-1}$, because $B = 1$.

Now consider the case when $B = 2$ and assume all component binary codes to be rate $R$ capacity achieving binary codes. Bit level reordering, encoding with component codes and mapping with $f_w(.)$ is shown in the Figure 1 and Figure 2. The rate of fall of distortion is $(1 + SNR)^{-2}$, because $B = 2$, while the bandwidth expansion factor is $N = 2/R$.

Consider the $i^{th}$ level binary encoded sequence $\mathbf{X}_i$. This sequence of coded bits suffers two independent additive noise components. One is the uniformly distributed noise due to the lower levels, while the other is the AWGN offered by the channel. Let us define $a_i \triangleq \frac{\sqrt{3}}{2^i}$. Then the coded bits are transmitted at $\pm a_i$, uniform noise is distributed in the interval $[-a_i, a_i)$ and the AWGN is $\mathcal{N}(0, \sigma^2)$. The probability of an encoded bit in the $i^{th}$ level being in error can be calculated as follows:

The combined additive noise pdf due to the uniform pdf and the Gaussian is given by:

$$f_z(\mathbf{z}) = \int_{\tau=-a_i}^{a_i} \frac{1}{2a_i\sigma\sqrt{2\pi}} e^{-\frac{(z-\tau)^2}{2\sigma^2}} d\tau = \text{erf}(\frac{z + a_i}{\sigma\sqrt{2}}) - \text{erf}(\frac{z - a_i}{\sigma\sqrt{2}})$$

which is the pdf of a random variable distributed as $\mathcal{GU}(a_i, \sigma^2)$.

Assuming equally likely encoded $\pm a_i$, the probability of an encoded bit in level $i$ being in error is given by,

$$P(i,\sigma) = \int_{z=0}^{\infty} f_z(\mathbf{z})dz = \frac{1}{2a_i\sigma\sqrt{2\pi}} \int_{-a_i}^{a_i} \int_{a_i}^{\infty} e^{-\frac{(x+t)^2}{2\sigma^2}} dxdt$$
$$= \frac{\sigma(1-e^{-\frac{2a_i^2}{\sigma^2}})}{2a_i\sqrt{2\pi}} + \frac{1}{2}\text{erfc}(\frac{a_i\sqrt{2}}{\sigma})$$

We can observe that for large SNR, coded bit error probability is dominated by $\sigma$, while for small SNR, it is dominated by the factor $(1-e^{-2a_i/\sigma^2})$. It is also seen that the probability depends only on the relative SNR of the $i^{th}$ encoded bit level, which is $a_i^2/\sigma^2$.

Now we consider the case when a systematic binary linear code with parameters $[n,k,d]$ is used as the component code. The probability of information bit error under hard-decision ML decoding when the code is used over a BSC with crossover probability of $p$ is upper bounded by:

$$P_e \leqslant \sum_{m=\frac{d-1}{2}+1}^{n} \frac{m}{n}\binom{n}{m} p^m(1-p)^{n-m} \tag{2}$$

Using the expression for $P(i,\sigma)$ for the crossover probability $p$, we get, for the $i^{th}$ level, the information bit error is upper bounded as:

$$P_e(i,\sigma) \leqslant \sum_{m=\frac{d-1}{2}+1}^{n} \frac{m}{n}\binom{n}{m} P(i,\sigma)^m(1-P(i,\sigma))^{n-m}$$

The average distortion due to the $i^{th}$ level is given by,

$$D(i,\sigma) = 2\sum_{j=0}^{B-1} a_{(Bi-j)}^2 P_e(i,\sigma) = 2\sum_{j=0}^{B-1} (\frac{\sqrt{3}}{2^{(Bi-j)}})^2 P_e(i,\sigma) = \frac{2(4^B-1)}{4^{Bi}} P_e(i,\sigma)$$

the summation above over $j$ was because errors in the $i^{th}$ coded bit level cause errors in the source bits planes $\{(Bi-j) : j \in \{0,\ldots,(B-1)\}\}$.

The distortion suffered at each level are independent and add up to the total distortion. If at the receiver, the analog decoder is limited to decoding the first $I$ levels, then the corresponding distortion is therefore:

$$D_I(\sigma) = 2^{-2BI} + \sum_{i=1}^{I} D(i,\sigma) = 4^{-BI} + 2(4^B-1)\sum_{i=1}^{I} 4^{-Bi} P_e(i,\sigma)$$

where the first term is the total distortion due to all source bit planes below level $BI$.

The performance of the analog codes constructed here can actually be better than the bound derived above if we use soft decision maximum likelihood decoders for the individual component codes. In practice, we can replace unbounded length capacity achieving codes with very good algebraic codes or with graphical codes. In the second case we have a low complexity near ML decoding algorithm readily available in the form of the message passing algorithm. Also, we can truncate the number of source bit planes, $I$ which are actually encoded to a manageable, yet reasonably high value.

## VI. Discussion and Examples

In this section, we examine two practical encoders and their performance to demonstrate the effectiveness of the new analog code construction, especially at higher SNRs.

### A. *Perfect Binary Golay Code* $[23, 12, 7]$ *as Component Code*

The rate of this code is $R = 23/12$. Since the code is perfect, in the bound of (2), equality holds.

We consider the case with $B = 2$. The bandwidth expansion factor, $N = \frac{Bn}{k} = \frac{46}{12} \approx 3.833$. In Figure 9, the lower bound on the distortion at a bandwidth expansion factor of $N \approx 3.833$ and $N = B = 2$ are shown, along with the upper-bound on distortion when this particular choice of component code is used along with hard decision ML decoding for the component codes at the analog decoder. We can see that, even with the number of source bit planes restricted to $I = 50$, we already see a rate of decline of distortion with increasing SNR which is proportional to the $N = B = 2$ case.

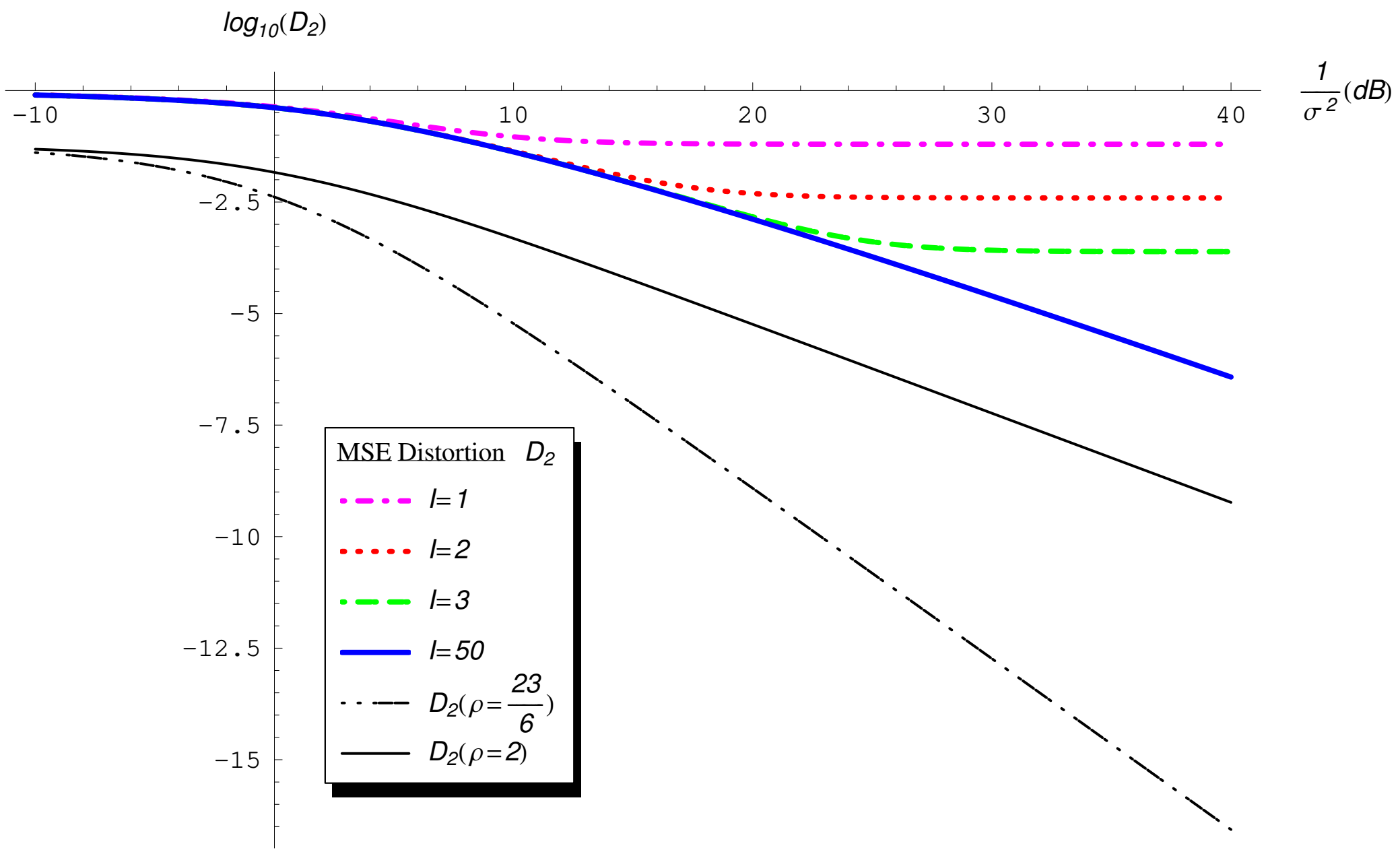

**Figure 9** *Upper bounds on the mean squared distortion of the proposed analog code constructed using the binary Golay code* $[23, 12, 7]$ *as the component code and hard decision ML decoders for the component codes at the receiver.* $\rho$ *denotes the bandwidth expansion factor.*

### B. *Binary Code* $[72, 36, 16]$ *as Component Code*

This code is taken from [MS77]. When $B = 3$, the bandwidth expansion factor is given by $N = B/R = 6$. From Figure 10, we see that the rate of fall in distortion is proportional to the $N = B = 3$ case.

For both the codes considered in this section, the actual performance can again be improved if we use soft decision decoders for the component codes.

### C. *Repeat Accumulate Codes as Component Codes*

As another verification of the code construction and decoding procedure, we employ the fairly powerful Generalized Repeat Accumulate $(RA^m)$ code [DJM98], with $m = 2$ as the binary code to construct an analog code. The random permuters used were of length 27000. This code choice enabled the use of a factor graph based Sum-Product Algorithm(SPA) [KFL01] at the decoder. But note that any other good graph based codes may also be used. Knowing the pdf $p_{\mathbf{z}_{\gamma,3/4}}(z)$, the messages from the leaf nodes of the Factor Graph (see Figure 11, Figure 13 and Figure 12) can be calculated and the message passing algorithm now proceeds through a forward backward schedule, which is terminated at a previously fixed iteration level. We simulated the sum-product algorithm for 20 iterations to report the curves of Figure 14. The simulation results are seen to be in excellent agreement with Theorem IV.1. The slope of the $D_2$(dB) - $\gamma$(dB) curve was $-B$ throughout the $SNR$ regime we simulated. The actual simulation curve is worse than the predicted performance from Theorem IV.1, by an SNR(dB) value by which the decoding threshold of the $(RA^2)$ code under iterative decoding differs from the Shannon limit for that channel. The decoder complexity increases linearly as the number of bits of precision, $M$ demanded at the receiver end. In a broadcast scenario, the transmitter transmits the same code symbols irrespective of the receiver SNR. Since the receiver knows the SNR, it can decode using the SPA at a resolution which can be supported by the code at the receiver SNR. Thus, in principle, the receiver can achieve any of the MSE distortion points predicted by Theorem IV.1. In practice however, the receiver performance is dictated by the performance of the binary component code. In particular, at higher SNRs, error floors of the binary codes can result in additional distortion.

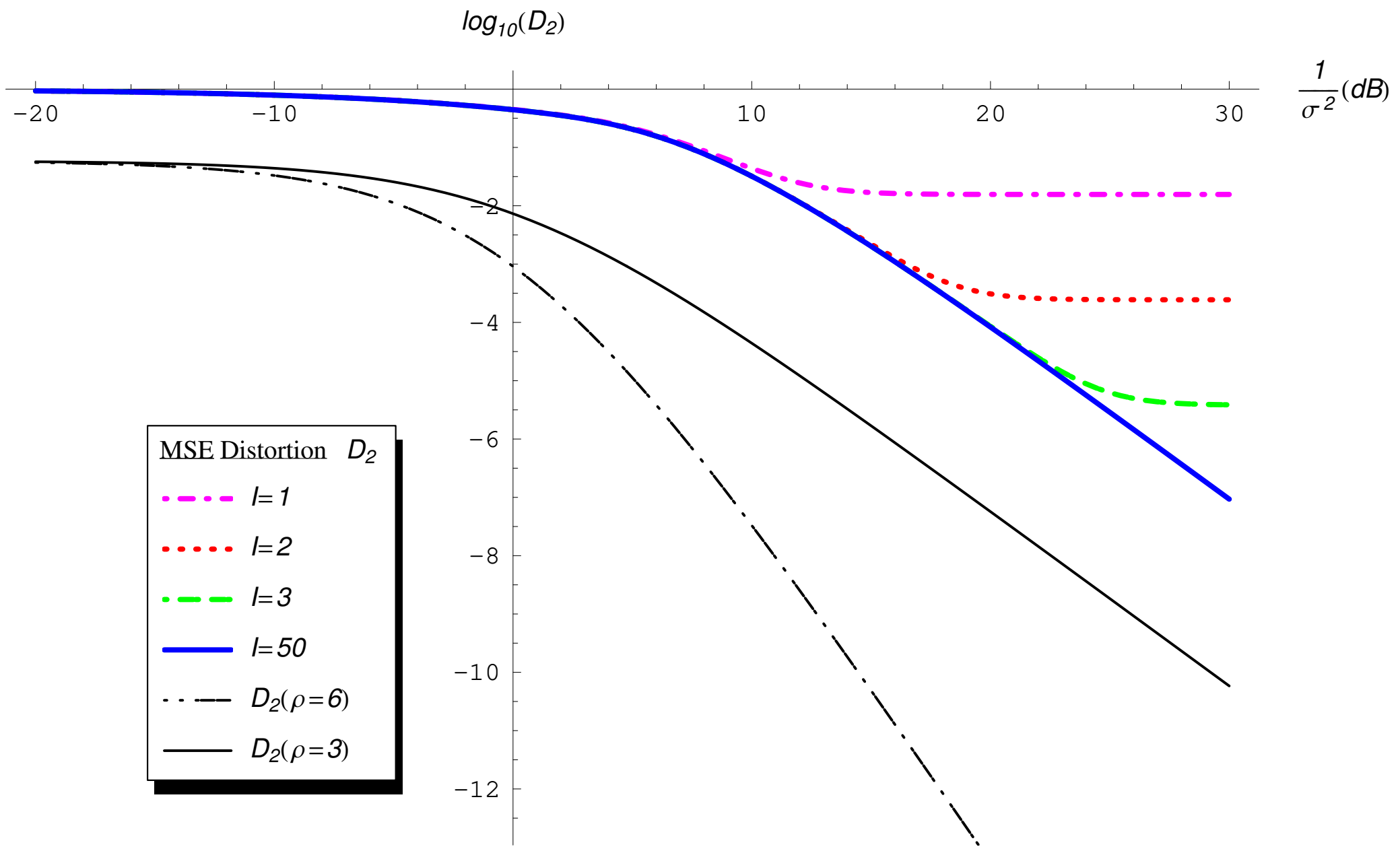

**Figure 10** *Upper bounds on the mean squared distortion of the proposed analog code constructed using the binary code $[72, 36, 16]$ as the component code and hard decision ML decoders for the component codes at the receiver. $\rho$ denotes the bandwidth expansion factor.*

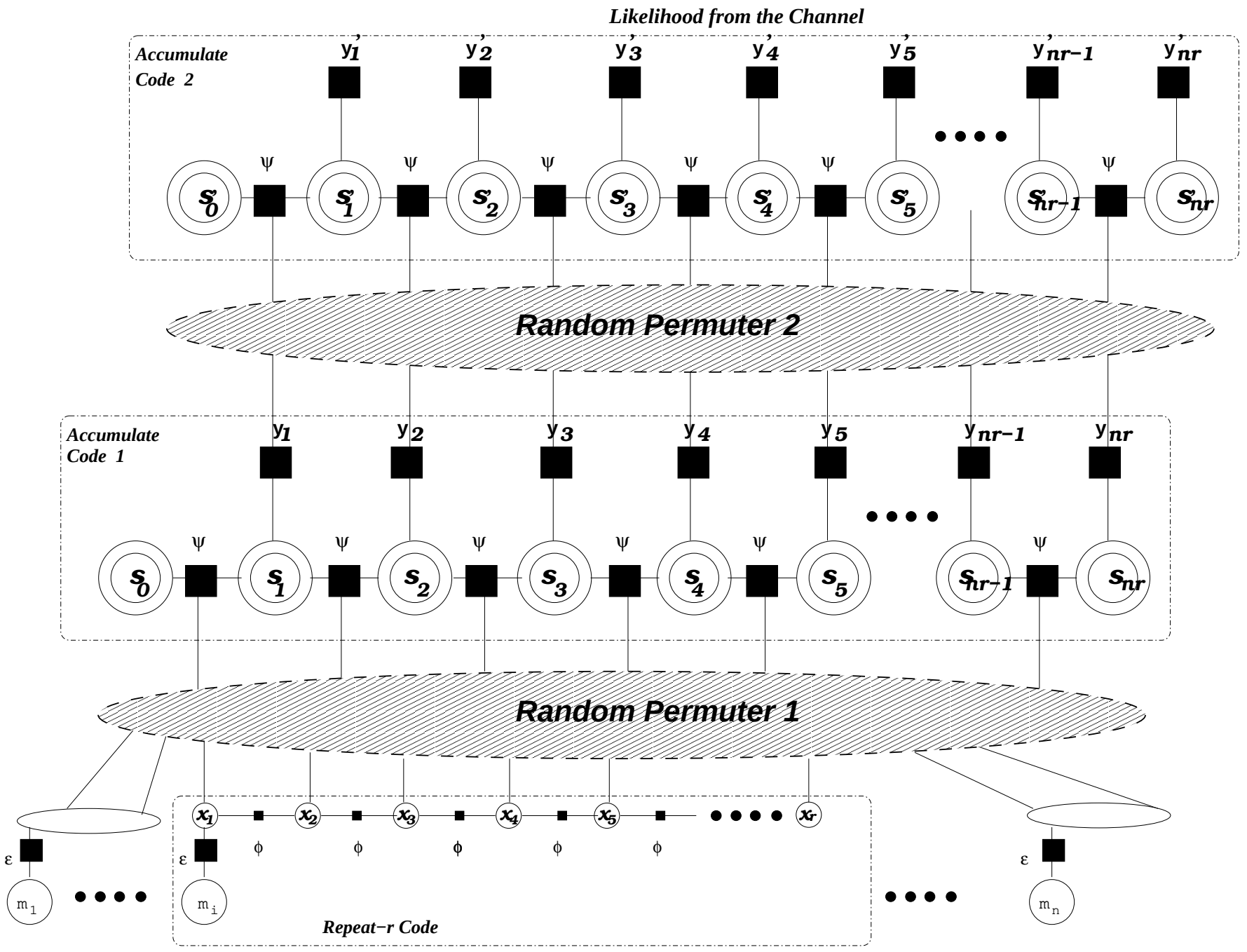

**Figure 11** *The factor graph for the $(RA^2)$ code.*

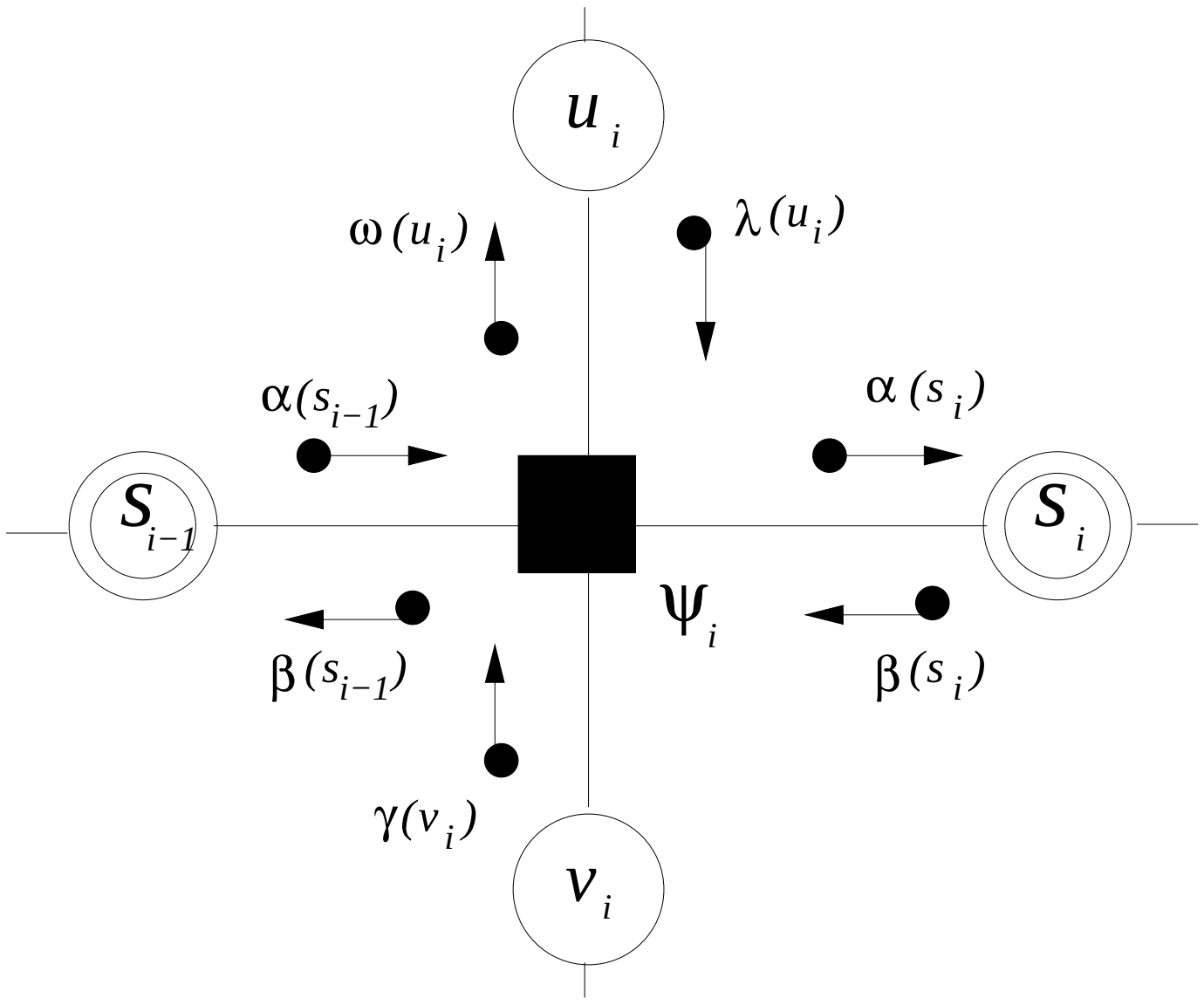

**Figure 12** *The messages into and out of a check (function) node.*

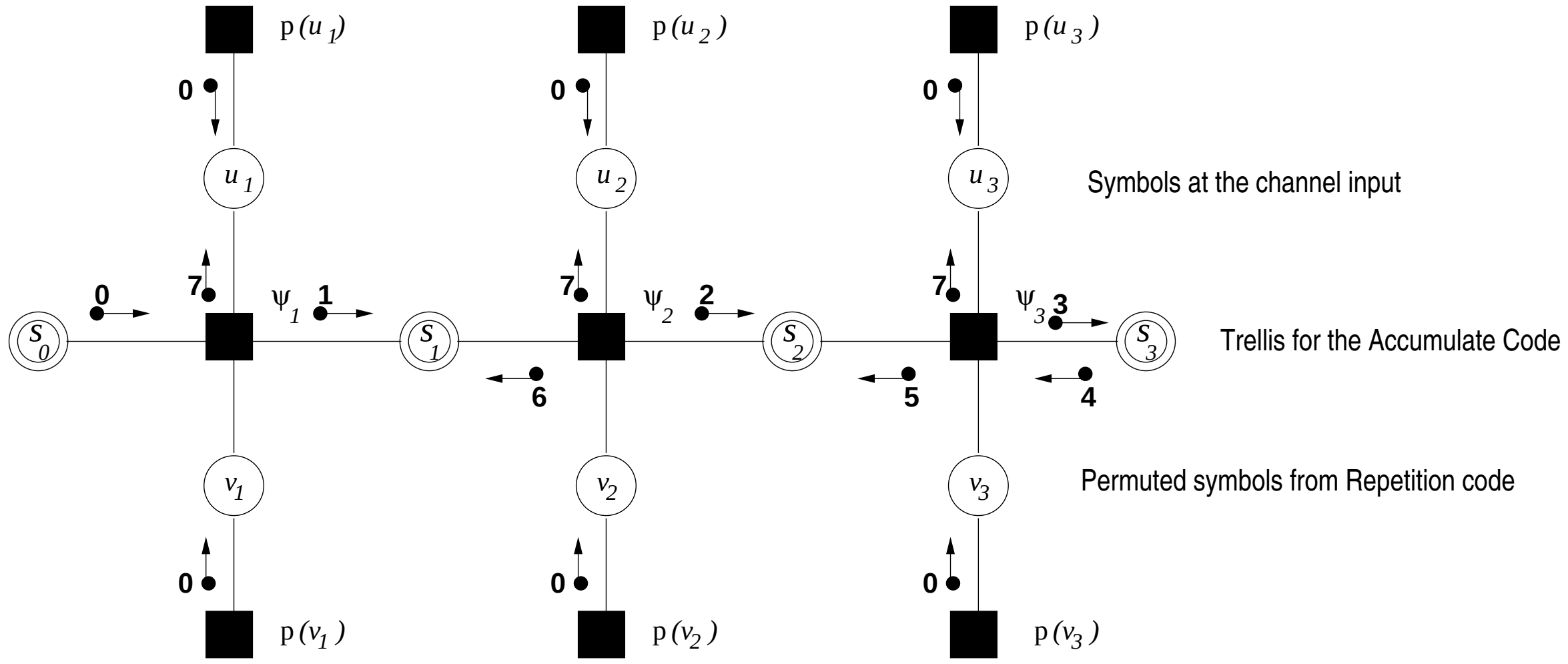

**Figure 13** *Forward backward algorithm on the Accumulate code graph. Here, $p(u)$ stands for the conditional pdf offered by the channel for the output, given an input symbol. Hence, for the Analog Codes, $p(u) = p_{\mathbf{z}_{\gamma,w}|y_{i,j}}(z|y)$, which is known in closed form for $w = 3/4$. The variables have a binary alphabet. For the truncated Analog Codes, $p(u)$ is a conditional Gaussian pdf. The variables in this case are from a $2^M$-ary alphabet.*

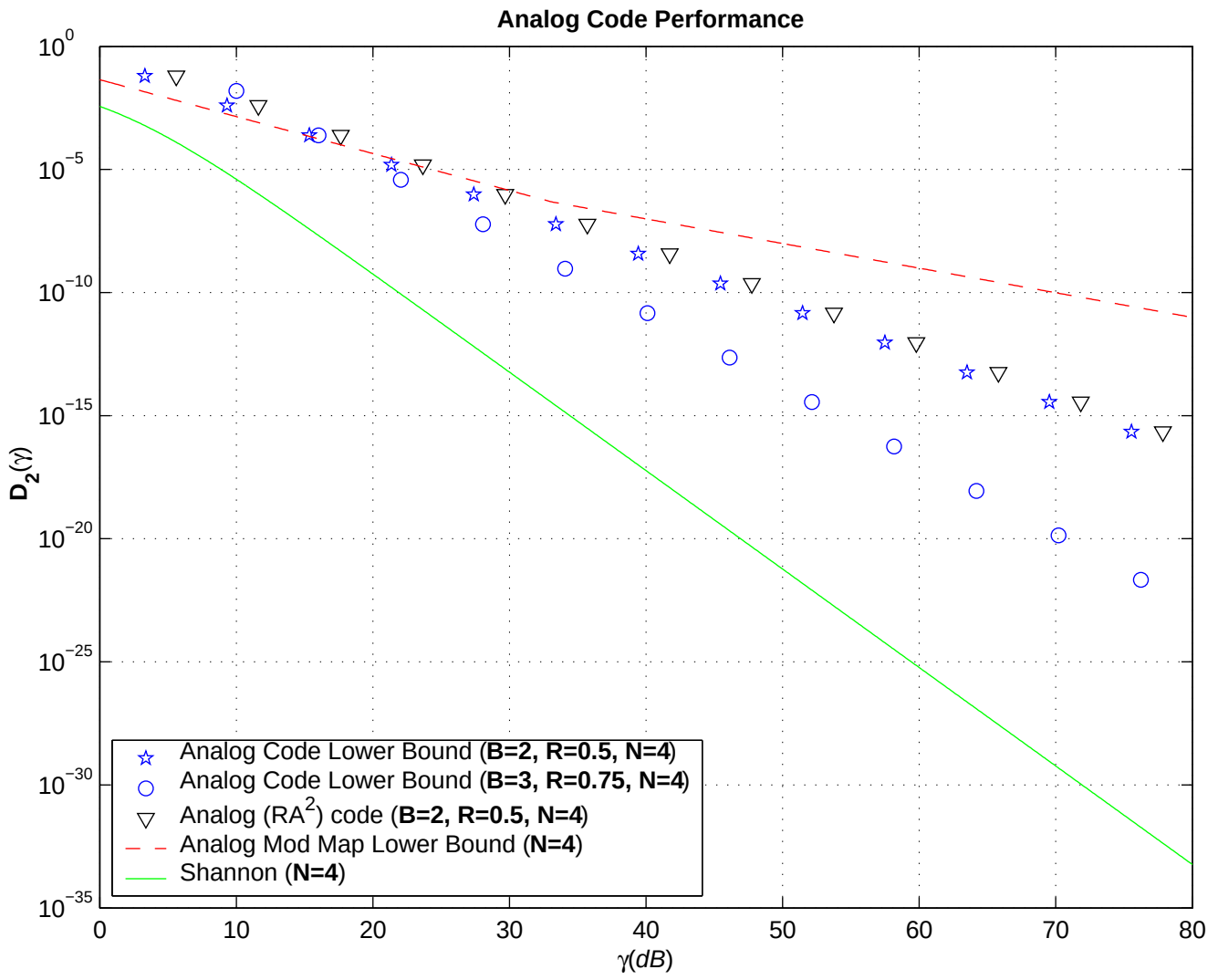

**Figure 14** *The code performance for bandwidth expansion $N = 4$ on AWGN channel SNR. Analog ($RA^2$) codes for $(B = 2, R = 0.5)$ and $(B = 3, R = 0.75)$ are plotted along with corresponding lower bound for the chaos codes of [CW98], and the Shannon lower bound, $D_2 \geqslant \frac{1}{2\pi e}(1 + SNR)^{-N}$ for the AWGN channel. The interleaver was of length 27000.*

### *D. Bandwidth efficient communication with truncated Analog Codes*

The effectiveness of analog codes based on capacity achieving component binary codes, prompts one to consider truncated versions of these for bandwidth efficient communication. The modulation codes that we consider first are analog graphical codes truncated to the $M$ most significant bits. Then for a code rate of $R$, the bits per channel use is seen to be $2M \cdot R$ for a complex AWGN channel. As a typical example, we simulated with generalized $(RA^2)$ code as the component codes for the analog code with $B = 1$ over a complex AWGN channel. The channel likelihood messages (see Figure 12, Figure 13 and Figure 11) are easily determined, since the channel conditional pdf is known to be Gaussian distributed. At the decoder, SPA is again employed (Figure 13). The code alphabet now has $2^M$ symbols. Hence, each message vector is of dimension $2^M$. At the end of the stipulated number of iterations, the source symbol with the largest associated áposteriori probability is selected.

A variation of the truncated analog codes has been found to perform better for small values of $M$ (at least for $M < 4$). Here, the Repeat Accumulate codes work on the ring $Z_{2^M}$. When $M = 1$, the resulting codes are the same as the binary $(RA^m)$ codes. The mapper converts the symbols in $Z_{2^M}$ to real numbers, using the natural mapping, scaled to achieve the desired average transmit power. The decoder uses the SPA( Figure 13). Now the component code trellises for the serial concatenation scheme draw from an alphabet with $2^M$ symbols. Again, for a complex AWGN channel, the channel conditional pdf is known to be Gaussian distributed, and at the end of the stipulated number of iterations, the source symbol with the largest associated áposteriori probability is selected. The results have been plotted in Figure 15, along with the computational cutoff rates of MPSK over complex AWGN channel.

## VII. Comparison with Prior Bounds on Distortion

In this section, we compare the new codes with the class of signals considered by Ziv. We show that the key condition of boundedness of “stretch-factor” does not hold for the codes constructed in this paper.

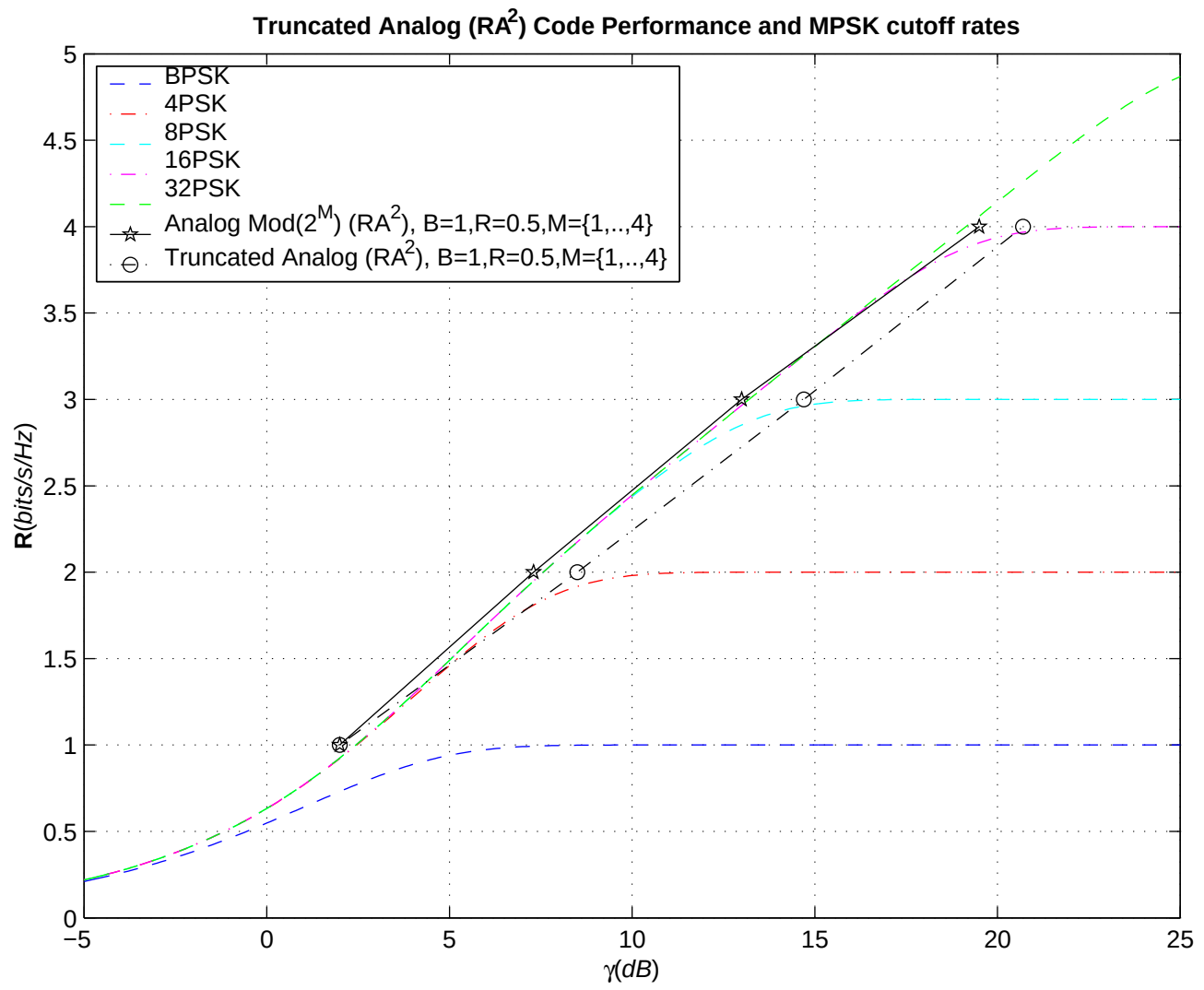

**Figure 15** *The performance of (a)Truncated Analog* $(RA^2)$ *codes for* $(B = 1, R = 0.5)$ *and* $M \in \{1,2,3,4\}$ *and (b)*$\mathbb{Z}_{2^M}(RA^2)$ *codes for* $M \in \{1,2,3,4\}$*, at* $BER = 10^{-5}$ *are plotted along with corresponding computational cutoff rate for MPSK on a complex AWGN channel. The interleaver was of length 27000.*

Therefore the threshold effect predicted in [Ziv70] does not necessarily apply to the new codes.

In [Ziv70] Ziv considers an analog data source which produces a sequence of real samples, $s_1, s_2, \ldots$ at a rate of $2W_S$ samples per second. The encoder maps a block of $k = 2TW_S$ such source samples, $S$, using a function, $f(t,S)$ such that the following hold:

$$[f(t,S)]^2 \leqslant f_{max}^2 < \infty$$
$$\frac{1}{T}\int_0^T E_S[(f(t,S))^2]dt \leqslant 1$$

If we restrict to the uniform analog source considered earlier, then Theorem 2 in [Ziv70] states that given a mapping $f(t,S)$ and positive integers $M, \alpha, \Delta_0$ such that for every $j = 1,2,\ldots,k$

$$d_j(\Delta) \stackrel{def}{=} \int_{-a}^{a}\int_{S:S_i=s_i}\int_0^T \frac{|\eth f_j(t,S,\Delta)|^2}{2a} ds_i dP(S|s_i)dt \leqslant M\Delta^\alpha \tag{3}$$

where

$$\eth f_j(t,S,\Delta) \stackrel{def}{=} f(t,s_1,s_2,\ldots,s_j+\Delta,\ldots,x_k) - f(t,s_1,s_2,\ldots,s_j,\ldots,x_k)$$

for any $\Delta \leqslant \Delta_0 \leqslant 2a$, then for any sufficiently large SNR $\gamma$, and a positive real number $K$,

$$D \geqslant \frac{K(2a-\Delta_0)}{(M\gamma)^{-2/\alpha}}$$

The function $d_j(\Delta)$ is called the *expected stretch factor*, and is assumed to be bounded by a number proportional to a positive power of $\Delta$. This assumption is valid for most practical analog communication systems such as FM, where there is a bandwidth expansion factor. However, for the analog codes constructed in this paper, the stretch factor is not polynomially bounded as a function of $\Delta$. We will show this by means of an example. Assume that all of the component codes are the binary $[7,3,4]$ code, which is the dual of the $[7,4,3]$ Hamming code. This code has the generator matrix given by:

$$\begin{bmatrix} 1 & 0 & 1 & 0 & 1 & 0 & 1 \\ 0 & 1 & 0 & 0 & 1 & 1 & 1 \\ 0 & 0 & 1 & 1 & 1 & 1 & 0 \end{bmatrix}$$

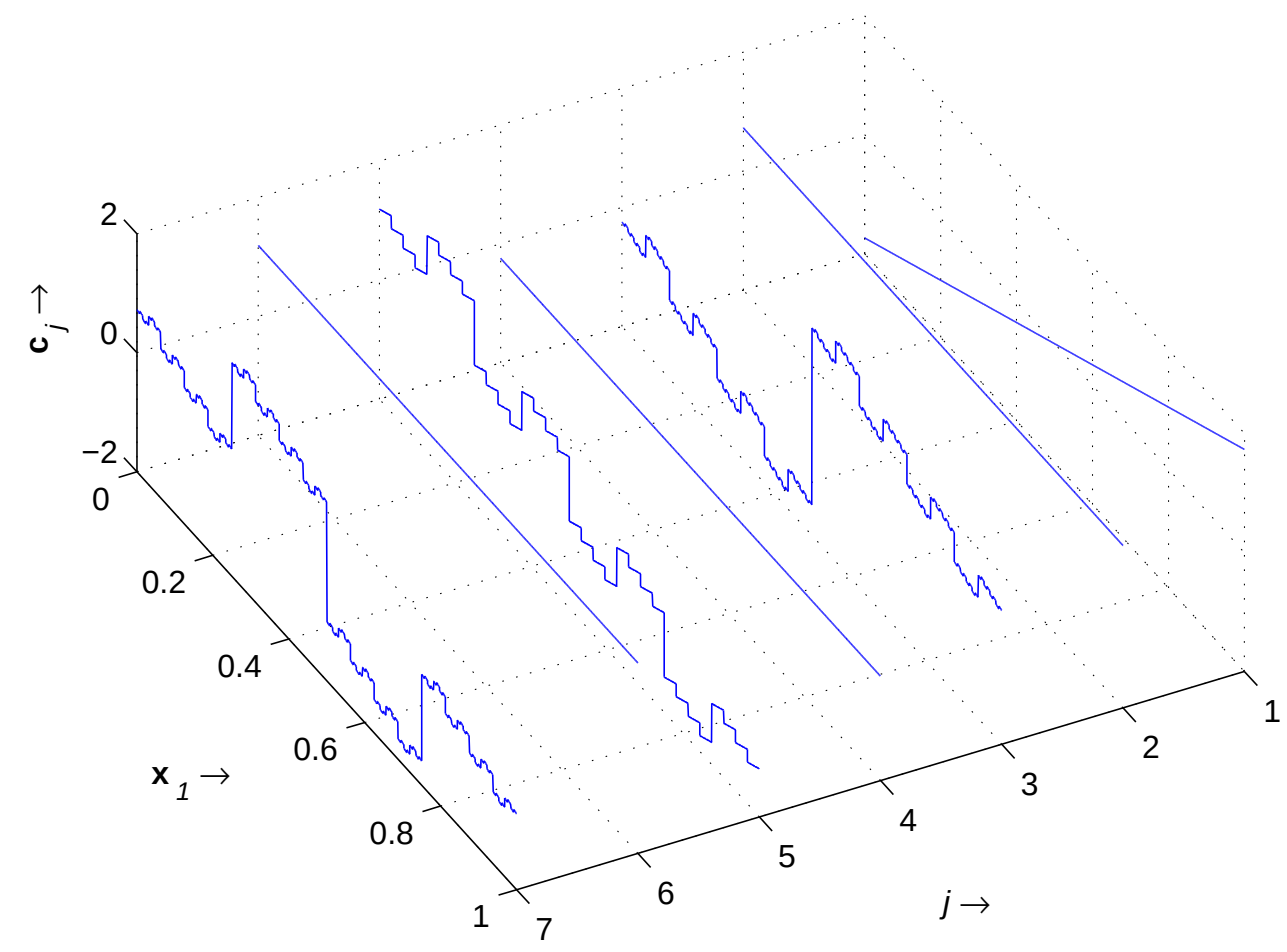

**Figure 16** *The dual Hamming code* $[7, 3, 4]$ *is used as a component code. The analog code sequences are shown when* $x_1$ *alone is changing, while* $x_2 = 0.7095$ *and* $x_3 = 0.4289$

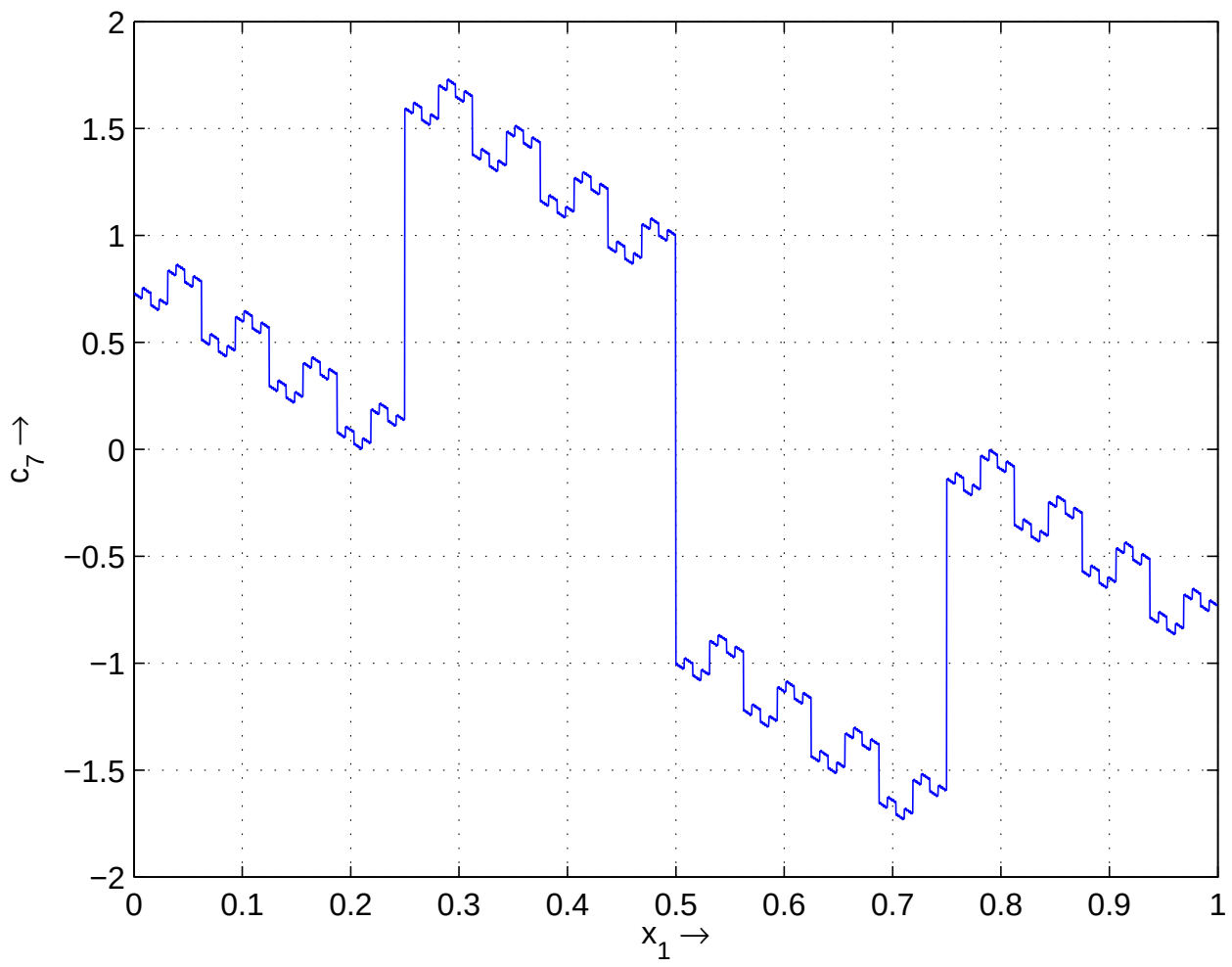

**Figure 17** *The dual Hamming code* $[7, 3, 4]$ *is used as a component code. Position number* 7 *in the analog code sequence is shown when* $x_1$ *alone is changing, while* $x_2 = 0.7095$ *and* $x_3 = 0.4289$

Figure 16 shows the analog code mapping as a function of $x_1$, while $x_2$ is held fixed at 0.7095 and $x_3$ is fixed at 0.4289. Recall that $x_j \triangleq (s_j + \sqrt{3})/2\sqrt{3}$, where $s_j$ are the samples from a band-limited analog source with a zero mean, unit variance uniform pdf. A very interesting situation is immediately apparent. For example, compare the map when $x_1 = 0.5 - \Delta$ and $x_1 = 0.5$. No matter how small the $\Delta$, the entire mapping changes from one point to the other. This is because all the bits in the binary representation of 0.5 are different from the bits in the binary representation of $0.5 - \Delta$, forcing all the component codewords to change. The same situation recurs when $x_1$ approaches all powers of $1/2$. In fact, the mapping is almost everywhere discontinuous as a function of $x_1$. This is an example of curve which is self-similar under scaling and shifting. Such forms are well known as fractals [Man82]. Similar observations can be found valid on all of the other three source dimensions as well.

Therefore, it is clear that the analog codes constructed in this paper do not fall in the class of modulation signals considered in [Ziv70].

One can also observe that, the proposed analog codes do not touch the corresponding rate-distortion based lower bounds on MSE distortion. Therefore, the conclusions similar to those of Reznic et al of [RFZ05] do not apply as well. Moreover, in [RFZ05], they consider Gaussian sources, whereas our construction is for a uniform source, though it may be possible to extend it to any source which admits successive refinement of information [EC91].

## VIII. Summary

We presented a new analog coding scheme, which can achieve a mean-squared error distortion proportional to $(1 + SNR)^{-B}$ for a bandwidth expansion factor of $B/R$, where $0 < R < 1$ is the rate of individual component binary codes used in the construction. Thus over a wide range of SNR values, the newly proposed code performs much better than any single previously known analog coding system.

## IX. Acknowledgments

Portions of this paper appeared in "Analog codes on graphs," *Proceedings of the International Symposium on Information Theory (ISIT)*, Yokohama, Japan, July 2003. Funding for this work was provided in part by research grants from the National Science Foundation. Material in this article is being prepared for a journal submission.